\documentclass[3p,times,twocolumn]{elsarticle}

\usepackage{ecrc}

\volume{00}

\firstpage{1}

\journalname{Nuclear and Particle Physics Proceedings}

\runauth{R Kunnawalkam Elayavalli and K C Zapp}

\jid{nppp}

\jnltitlelogo{Nuclear and Particle Physics Proceedings}

\usepackage{amssymb}

\usepackage[figuresright]{rotating}

\begin{document}

\begin{frontmatter}



\dochead{}

\title{Medium Recoils and background subtraction in JEWEL}


\author[label1]{Raghav Kunnawalkam Elayavalli}
\ead{raghav.k.e at SPAMNOT cern.ch}
\author[label2,label3,label4]{Korinna Christine Zapp}
\address[label1]{Rutgers, The State University of New Jersey, Piscataway, New Jersey 08854, USA}
\address[label2]{CENTRA, Instituto Superior T\'ecnico, Universidade de Lisboa, Av. Rovisco Pais, P-1049-001 Lisboa, Portugal}
\address[label3]{Physics Department, Theory Unit, CERN, CH-1211 Gen\`eve 23, Switzerland}
\address[label4]{Laborat\'{o}rio de Instrumenta\c{c}\~{a}o e F\'{\i}sica Experimental de Part\'{\i}culas (LIP), Av. Elias Garcia 14-1, 1000-149 Lisboa, Portuga}

\begin{abstract}
\textsc{Jewel} is a fully dynamical event generator for jet evolution in a dense QCD medium, which has been validated for multiple jet and jet-like observables. Jet constituents (partons) undergo collisions with thermal partons from the medium, leading to both elastic and radiative energy loss. The recoiling medium scattering centers carry away energy and momentum from the jet. Keeping track of these recoils is essential for the description of intra-jet observables. Since the thermal component of the recoils is part of the soft background activity, comparison with data on jet observables requires the implementation of a background subtraction procedure. We will show two independent procedures through which background subtraction can be performed and discuss the impact of the medium recoil on jet shape observables and jet-background correlations. Keeping track of the medium recoil significantly improves the \textsc{Jewel} description of jet shape measurements.
\end{abstract}


\end{frontmatter}

\section{Introduction}
	High energetic partons propagating through the quark gluon plasma (QGP) lose energy through elastic, inelastic and radiative processes. One way of studying the complex nature of energy loss is by analyzing the structure of reconstructed jets. Recent measurements of jet substructure in heavy ion collisions highlight several key features of the medium induced modifications. One of the measurements that generated significant discussions in the community is the sub-jet groomed momentum fraction from CMS~\cite{cmssplitting}. Due to the sensitivity of these measurements to medium-jet interactions, they have the potential of differentiate between different theory/MC calculations. In this report, we look at the recent developments to the \textsc{Jewel}~\cite{KZjewel1} heavy ion generator and how a careful treatment of the background and its subtraction, along with a estimation of the systematic uncertainties involved, leads to a better comparison with data and the physics that can be extracted from the results. 

\section{Medium recoils and background subtraction}
	\begin{figure}[h] 
	   \centering
	   \includegraphics[width=0.47\textwidth]{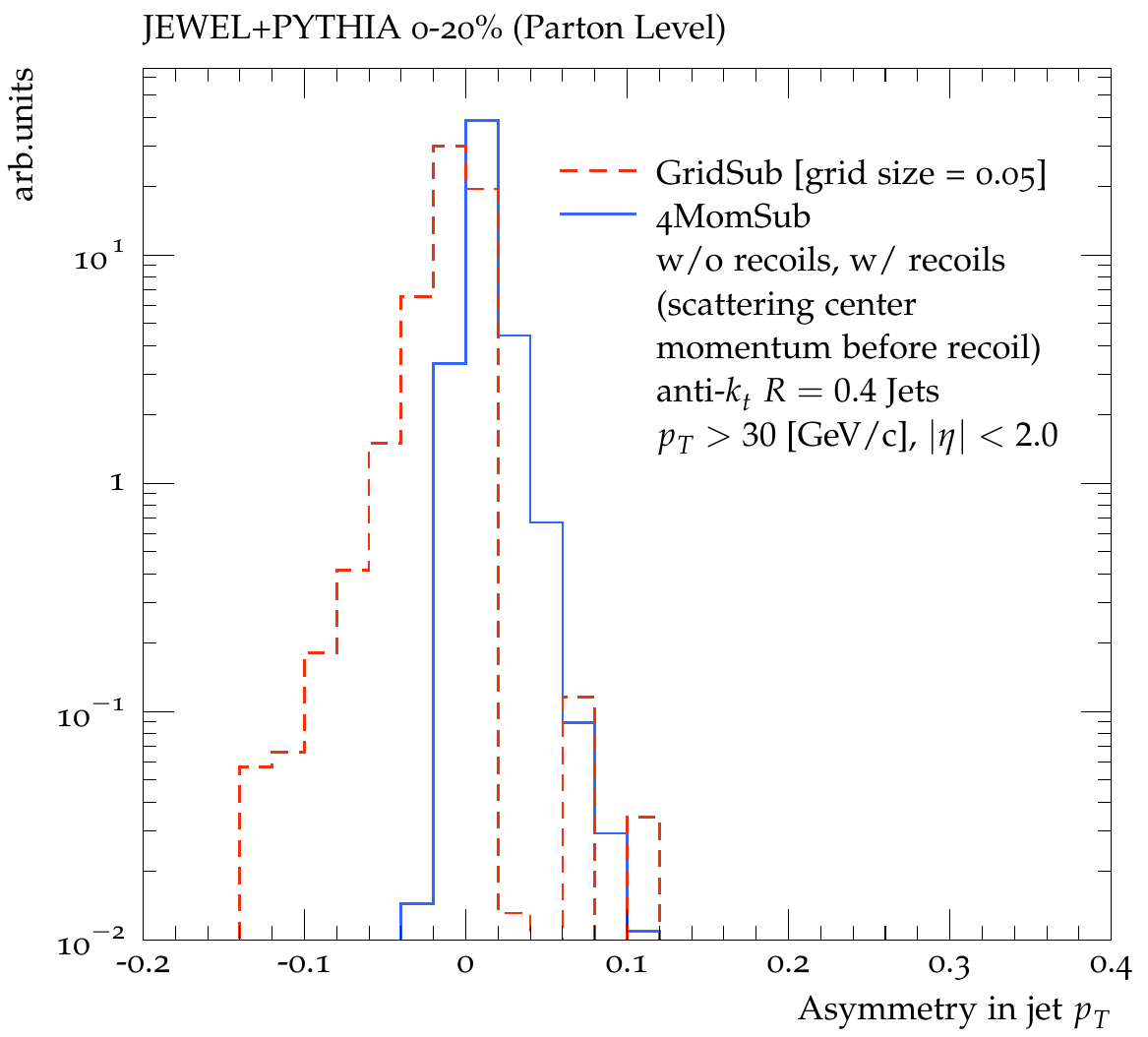} 
	   \caption{Asymmetry in parton level jet $p_{T}$ with removal of the recoils compared to background subtraction techniques. The GridSub, in dotted lines is more broader as opposed to the 4MomSub technique shown in solid lines.}
	   \label{fig:subAsym_partonLevel}
	\end{figure}		
	\begin{figure*}[h!] 
	   \centering
	   \includegraphics[width=0.47\textwidth]{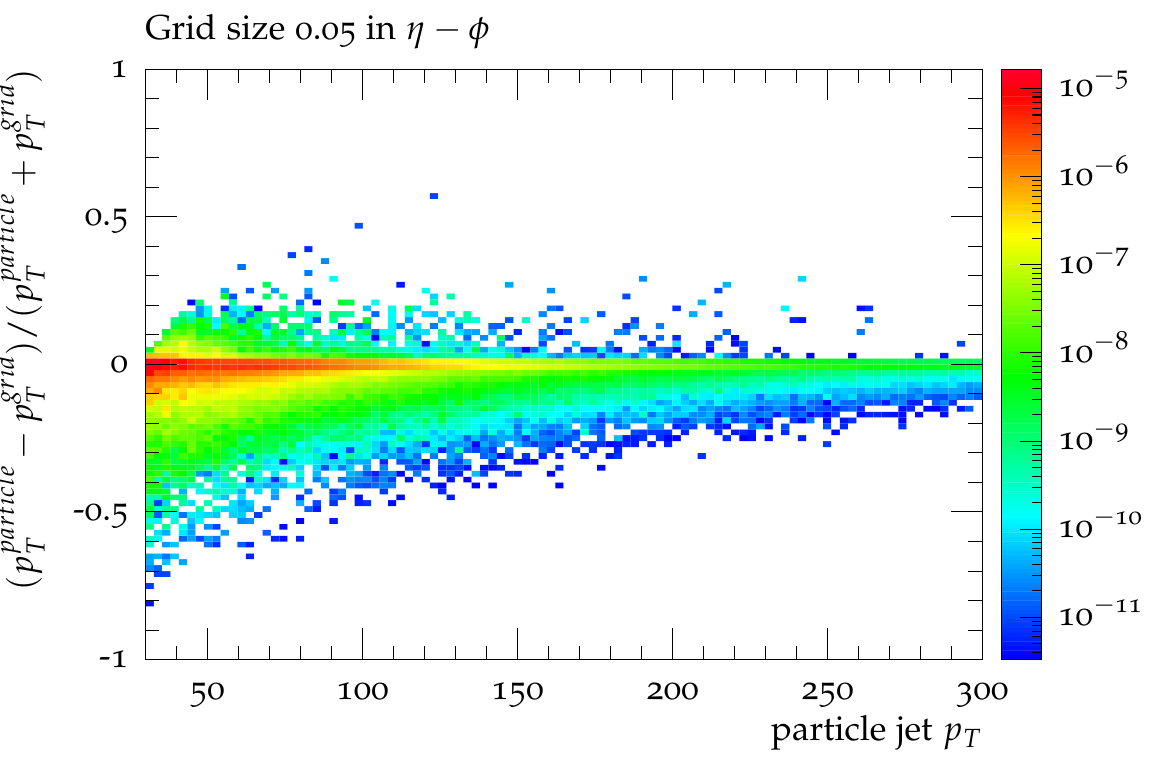} 
	   \includegraphics[width=0.47\textwidth]{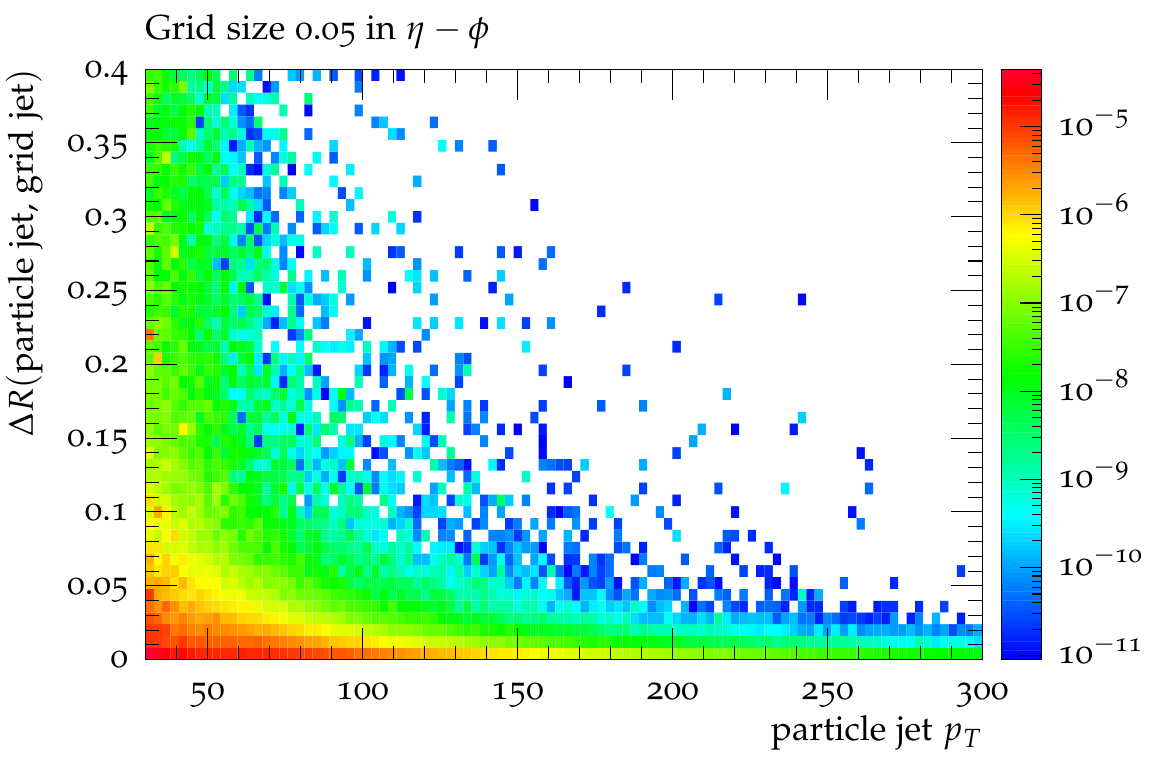} 
	   \caption{Smearing introduced by the grid on the particle level jet, quantized by the asymmetry in the jet $p_{T}$ on the left and the absolute shift of the jet axis in the $\eta-\phi$ plane, on the right, respectively shown as a function of the particle jet $p_{T}$. As expected, with increasing jet $p_{T}$, the effect of the grid becomes much smaller.}
	   \label{fig:gridjetsmearing}
	\end{figure*}	
	
	\textsc{Jewel} performs the final state medium modified jet evolution on dijet events generated by \textsc{Pythia6}~\cite{py6}. The implementation of energy loss in \textsc{Jewel} is through microscopic interactions, both elastic and inelastic with a thermal distribution of scattering centers along with a MC version of the LPM effect~\cite{KZjewel1,KZjewel2}. The partons that scatter upon interaction with these scattering centers are known as the recoiled partons. The recoils are a consequence of the medium interaction and once created, they do not undergo further interactions with other scattering centers. Hence one can run \textsc{Jewel} with two modes that correspond to keeping the record of the recoil partons leading to the final state or remove them before hadronization. For jet observables that only depend on overall jet quantities such as transverse momenta ($p_{T}$) or the jet axis, it is sufficient to run \textsc{Jewel} without storing the information of the recoil partons. On the other hand, for observables that depend on the structure of the jet such as the jet mass or profile etc. it is imperative that the recoil information is stored. Here a mismatch arises when comparing observables estimated using \textsc{Jewel} with recoils to data. This mismatch is a consequence of the systematic removal of the fluctuating heavy ion underlying event in data, whilst in \textsc{Jewel} the recoil partons still carry with them the thermal component of the scattering centers. It is precisely the momenta of the scattering center before interaction that needs to be subtracted and \textsc{Jewel} is now capable of storing that information as comments in the HepMC file. In order to remove this background component to the jets in \textsc{Jewel}, two background subtraction techniques are introduced. 
	\begin{itemize}
		\item 4MomSub: Neutral particles with very small momenta that are pointing back to the scattering centers are added to the final state particle list. These dummy particles get clustered to the jet but do not affect the jet itself. Then the dummy particles are matched to the scattering centers in position. The sum of the four-momenta of the matched scattering centers constitute the background of the jet and they are vectorially subtracted from the jet's four momenta. 
		\item GridSub: Similar to the previous method, we have two collections of objects comprising the final state particles (without the dummy particles in this case) and the scattering centers. A finite resolution grid is superimposed on the event and inside each cell, the four momenta of the scattering centers are subtracted from the final state particles that fall inside the grid. Then the jets are clustered using a single four momentum from each cell as input. If a cell turns out to have only scattering centers or if the total $p_{T}$ of the scattering centers exceeds that of the final state particles, that cell's four momenta is set to zero and it does not participate in the jet clustering. 
	\end{itemize} 

	These background subtraction methods were systematically studied for any biases or smearing they might have on the jet collection and the observable at interest~\cite{upcomingBackground}. The routines are implemented in the RIVET~\cite{rivet} analysis framework and jets are reconstructed using the FastJet~\cite{fastjet} toolkit with criteria similar to experimental measurements.  

\section{Systematic studies}
	
	As a first validation of the background subtraction, two sets of jets from the same events were compared. At the parton level, \textsc{Jewel} w/ recoils and background subtraction when compared to \textsc{Jewel} w/o recoils should yield similar jets if the background subtraction procedure is efficient. The asymmetry in the jet $p_{T}$ for both the background subtraction procedures is shown in Fig:~\ref{fig:subAsym_partonLevel} for mid rapidity jets with $p_{T} >30$ [GeV/c]. The red dashed lines represents the GridSub method which is seen to be under-subtraction precisely due to the fact that it doesn't take into account the cells that only contained the scattering centers, along with the smearing imparted to jet due to the finite resolution. The grid size is optimally chosen to be $0.05$ in $\eta,\phi$ and the systematics are estimated by varying that by a factor of two. The 4MomSub method is shown in the blue solid lines and its not a delta function at zero because the jet area can still increase due to the presence of additional objects in the jet cone. The deviations from zero becomes significantly less likely when the jet $p_{T}$ is increased. Similar to the detector resolution, GridSub method smears the jet's momenta and the axis, both of which are presented on the top and bottom of Fig:~\ref{fig:gridjetsmearing}. For high $p_{T}$ jets, the smearing decreases as expected. Due to the nature of the scattering centers being before hadronization, the subtraction procedure is only accurate for full jets. An accurate subtraction for charged jet (and or individual tracks etc.) are not in scope of the current version of \textsc{Jewel}.

\section{Predictions and comparisons with data}
	
\begin{figure}[h] 
	   \centering
	   \includegraphics[width=0.47\textwidth]{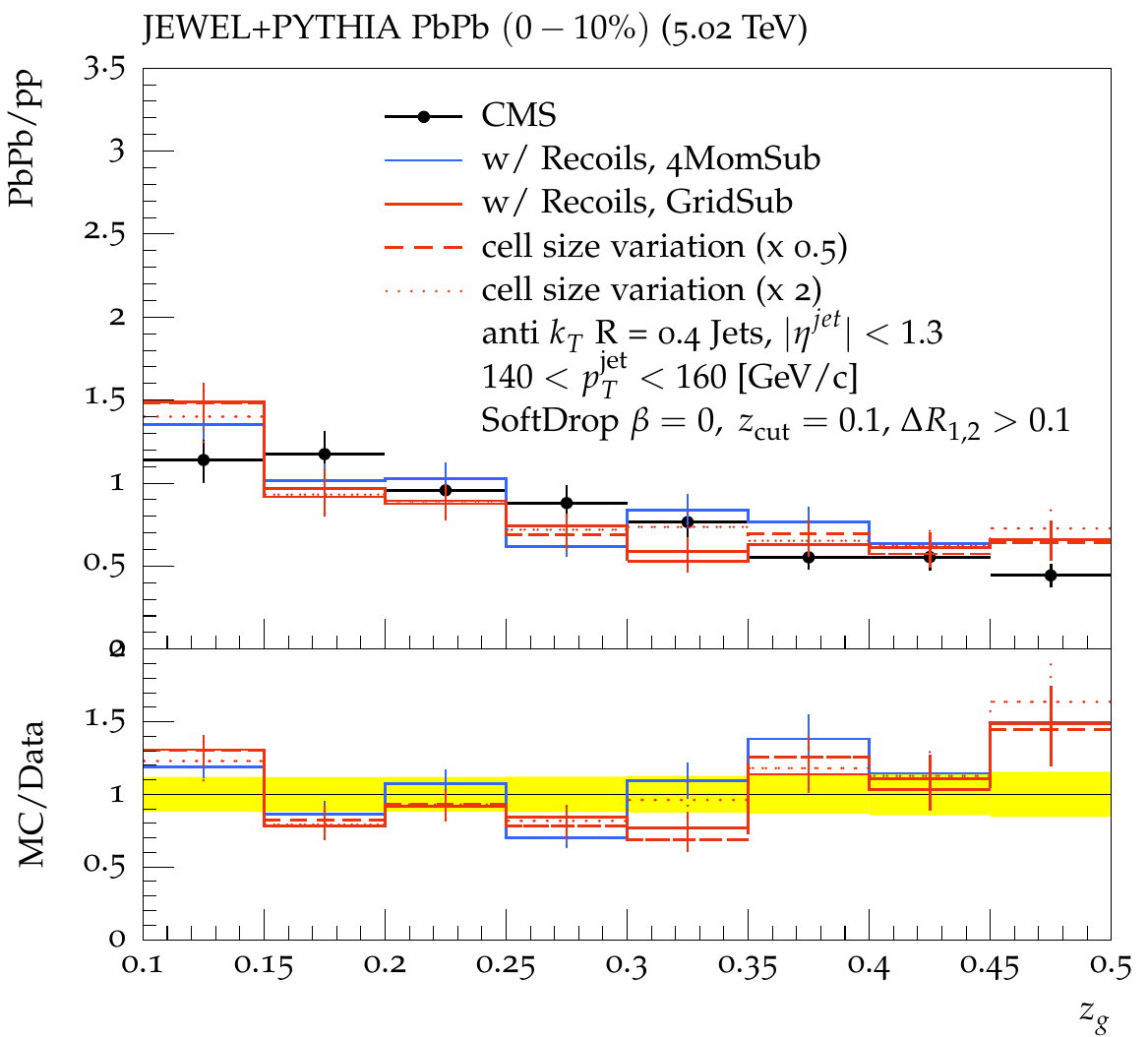} 
	   \caption{Comparison of \textsc{Jewel+Pythia} predictions to CMS data for the ratio of the subjet shared momentum fraction distributions in central PbPb events to pp events. The yellow shaded region around unity on the bottom panel highlights the data systematic uncertainties.}
	   \label{fig:cmsSplitting}
	\end{figure}

	The effect of the medium on jet structure such as its fragmentation or splittings, are multifold and hence enough care needs to be provided to the technicalities involved during any measurement of medium induced modifications.  Jet grooming tools are well studied and utilized in the high energy pp community in order to distinguish between signal and background for boosted objected resonance searches. Softdrop grooming~\cite{sdgrooming} is one of those tools that helps clean the soft component in a jet and leave the core structure untouched, leading to specially designed discriminants to isolate two prong structure from one prong structure. It does this by taking anti-k$_{t}$ clustered jets and re-clustering them using the CA algorithm. Thus by walking back through the jet clustering history on the CA jet, individual legs that do not pass the criteria are removed.  In heavy ion collisions, the soft component in a jet is an essential part of the medium-jet interactions and thus softdrop provides an handle, both experimental and theoretical, on isolating the response of the medium to the jet structure. CMS has recently measured the subjet groomed momentum fraction in PbPb jets compared to a pp baseline~\cite{cmssplitting}. The observable of interest here is ratio of the lower $p_{T}$ sub-jet in the jet to the sum of the two exclusive subjets after the grooming procedure; $$z_g = \frac{\rm{min}(p_{T,1}, p_{T,2})}{p_{T,1}+ p_{T,2}} > z_{\rm{cut}} \left(\frac{\theta_{1,2}}{R_{j}}\right)^{\beta}$$ where $\theta_{1,2}$ is the distance between the two subjets and $R_{j}$ is the jet radii. The distribution yields a degree of sensitivity to the first hardest splitting in the parton based on a threshold of $10\%$ denoted by $z_{\rm{cut}}$, and $\beta = 0$ denotes the absence of any constraint on the angular requirement during the grooming procedure. The ratio of $z_{g}$ in PbPb jets to pp jets is shown in Fig:~\ref{fig:cmsSplitting} where the CMS data is in black points and \textsc{Jewel+Pythia} predictions for the different background subtraction procedures are shown in the solid colored lines. The systematics are shown in dotted and dashed lines for the GridSub method as a size variation in the grid. The bottom panel shown the ratio of the prediction to data and \textsc{Jewel+Pythia} reproduces the general behavior that points to more asymmetrical parton splittings in PbPb as opposed to pp. The CMS preliminary result is not unfolded but the pp is smeared by the resolution of the PbPb. For this particular observable, due to the ratio, the smearing does not alter the final result (larger than the limit set by its systematic uncertainties) and can thus be compared directly with MC/Theory predictions.  
	
	\begin{figure}[h] 
	   \centering
	   \includegraphics[width=0.47\textwidth]{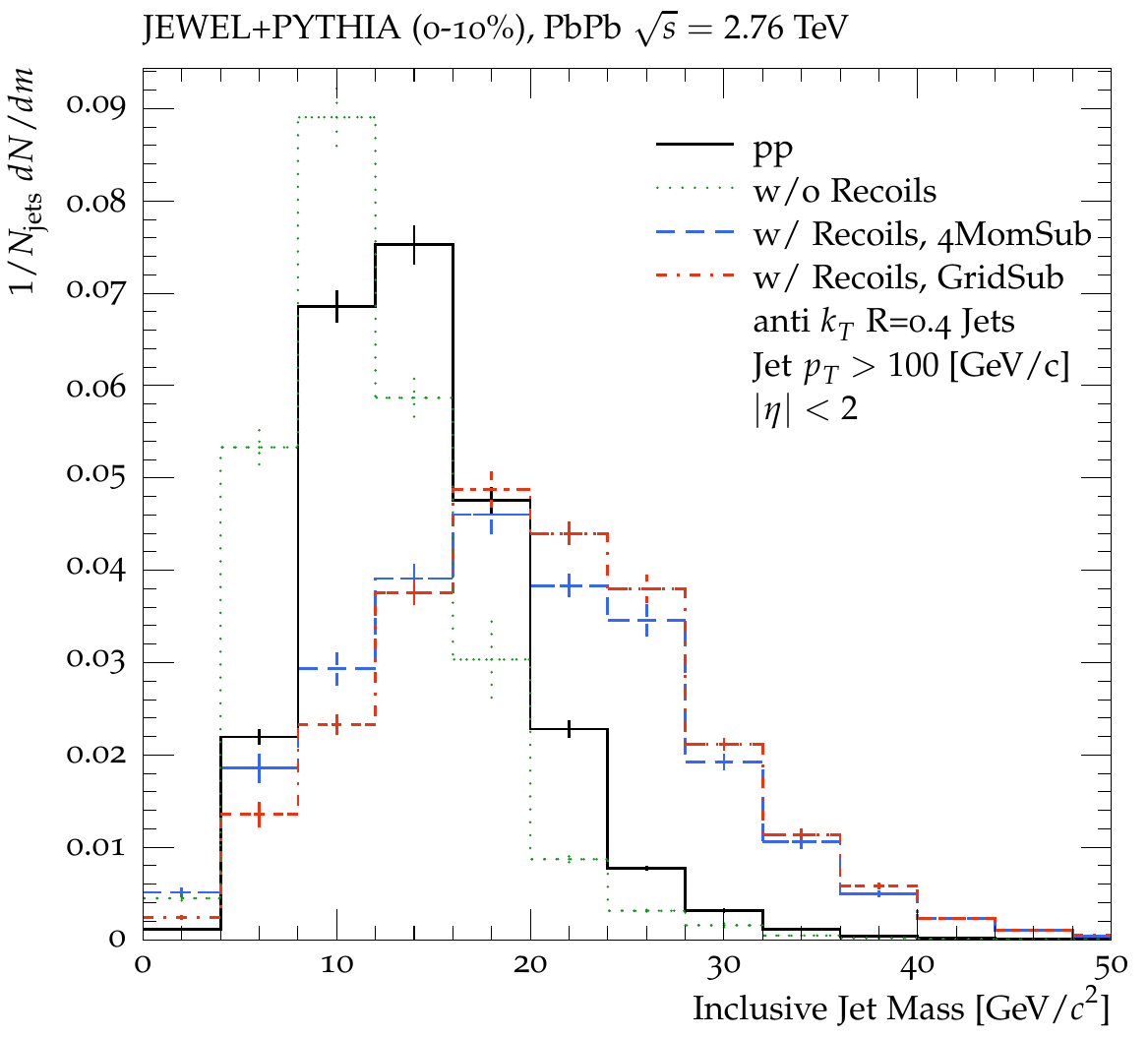} 
	   \caption{\textsc{Jewel} predictions for inclusive jet mass for pp (black solid line) and central PbPb with line color/styles representing the different method in the kinematic range. With recoils and background subtraction predicts a larger jet mass in PbPb when compared to pp collisions.}
	   \label{fig:aliceChJetMass}
	\end{figure}
	
	A measurement of the invariant jet mass in heavy ion collisions is very promising due to the sensitivity of the mass towards the soft sector in the jet cone. The motivation lies in the fact that earlier measurements such as the jet shapes/profile and the jet fragmentation function have shown the increased multiplicity of low $p_{T}$ particles at the periphery of the jet along with a narrowing of the jet cone. The predictions for full jets from \textsc{Jewel} are shown on the top of Fig:~\ref{fig:aliceChJetMass} for the different subtraction procedures and as one expects the jet mass is larger in PbPb with recoils when compared with pp. 

\section{Conclusions}
	Results on the modifications to the jet internal structure offer a unique view on the different jet energy loss models available today.  The treatment of the important medium-jet interactions in \textsc{Jewel} are envisioned by the usage of thermal scattering centers. This thermal component needs to be subtracted when observables are built using the recoils for which two new methods are presented. These methods, 4MomSub and GridSub are systematically studied and their corresponding effects on the jets are documented. After subtraction, \textsc{Jewel} is now able to match data for a plethora of jet shape and structure observables across kinematic ranges except the recent charged jet mass measurement for which \textsc{Jewel}, along with several other energy loss models over-predicts the mass. This is a very interesting time period where detailed studies on the jet structure are under way and these new upcoming results promises to provide further clues to solve the puzzle of jet energy loss and structure modifications due to the QGP.  
	
\section*{Acknowledgements}
This work was supported by Funda\c{c}\~{a}o para a Ci\^{e}ncia e a Tecnologia (Portugal) under project CERN/FIS-NUC/0049/2015 and postdoctoral fellowship \\SFRH/BPD/102844/2014 (KCZ) and by the European Union as part of the FP7 Marie Curie Initial Training Network MCnetITN (PITN-GA-2012-315877) (RKE). RKE also acknowledges support from the National Science Foundation under Grant No.1067907 \& 1352081. 
We also like to thank Dr. Chun Shen for providing the initial hydrodynamics parameters for our event generation.

\bibliographystyle{elsarticle-num}

\end{document}